# Effects of Si solute on the glass formation and atomic structure of Pd liquid

Z. J. Yang[1,2], L. Tang[1,2, *], T. Q. Wen[2,3], K. M. Ho[2,4], and C. Z. Wang[2,4, †]

[1]*Department of Applied Physics, College of Science, Zhejiang University of Technology, Hangzhou, 310023, China*

[2]*Ames Laboratory-USDOE, Iowa State University, Ames, Iowa 50011, USA*

[3]*MOE Key Laboratory of Materials Physics and Chemistry under Extraordinary Conditions, School of Natural and Applied Sciences, Northwestern Polytechnical University, Xi'an 710072, China*

[4]*Department of Physics and Astronomy, Iowa State University, Ames, Iowa 50011, USA*

Corresponding authors:   [*] lingtang@zjut.edu.cn       [†] wangcz@ameslab.gov

## Abstract

Molecular dynamics simulations were performed to study the effects of Si solute on the glass formation and crystallization of Pd liquid. Pure Pd sample prepared by quenching process with cooling rate of $10^{13}$ K/s can be in an amorphous state but the structural analysis indicates there is nearly no glass-forming motif in the sample. However, doping a small amount Si (Si concentration ~ 4%) the sample can be vitrified at a cooling rate of $10^{12}$ K/s. The glass-forming motifs such as Pd-centered Z13, Si-centered Z9-like and Mixed-ICO-Cube clusters with 5-fold local symmetry are found to be the dominant short-range orders in the glassy samples. With the increasing of the Si-doping concentration, these glass-forming motifs tend to aggregate and connect with each other forming a network structure. Our calculated results revealed that Si solutes in liquid Pd can significantly enhance the glass-forming ability.

I. Introduction

Since the first striking success in synthesis of metallic glass (MG) as early as 1960s [1], bulk MG has attracted a great deal of attentions due to its extraordinary mechanical properties [2-10]. Nowadays, one of the major challenges is to achieve MG state in monoatomic metals. Because the glass-forming ability (GFA) of monoatomic metals is extremely weak, the cooling rate required for forming bulk MG in such systems is usually much faster than that can be achieved by conventional experimental quenching methods. Several years ago, Zhong et al. [11] have developed an experimental technique to achieve ultrafast cooling rate up to $10^{14}$ K/s and successfully obtained metallic glass for tantalum.

On the other hand, high cooling rate is more suitable for molecular dynamics (MD) simulations. Cooling rate in common MD simulation is about $10^{10}$ K/s or higher. Hence, MD simulation can be applied to investigate the structure and glass formation of pure metals at ultrahigh cooling rates. Meanwhile, the atomic trajectories from the MD simulations can provide much more detail information and insights into the glass formation and crystal nucleation process, e.g., through the recently developed cluster alignment method [12], which is a very reliable tool to identify the short-range order (SRO) and medium-range order (MRO).

In recent years, a lot of experimental [13-18] and theoretical studies [19, 20] have been devoted to Pd and Pd-Si systems. Using the experimental approach of Ref. 11 discussed above , monatomic metal palladium is still unable to form glass state even under the ultrafast cooling rate of $10^{14}$K/s [11]. It is interesting to predict and estimate the cooling rate required for vitrifying pure Pd by MD simulations. On the other hand, it has been demonstrated that the binary $Pd_{80}Si_{20}$ alloy has an excellent GFA. Therefore, it is of great interest to investigate how a small amount of Si solutes can affect the glass formability in Pd.

Although the Pd-Si system with different composition of Si has been studied by some MD simulations [21, 22], detailed analysis of atomic structure of Pd under influence of small amount of Si solutes is still very limited. Recently, first-principles MD calculation [23] demonstrated that two types of Si-centered motifs dominate the SRO of $Pd_{82}Si_{18}$. One is a trigonal prism capped with three half-octahedra (Z9), and the other is archimedean anti-prism (Z10). These clusters tend to connect with each other and form a network, playing an important role in glass formation of $Pd_{82}Si_{18}$. For the Pd-Si system with very small Si concentration (i.e., atomic Si concentration from 2% to 5%), it is of interest to study the existence of these glass-forming Z-clusters and their spatial distribution, which could provide more insights into the solidification process in Pd-Si system.

In this paper, the atomic structure of pure Pd and Si-doped Pd with different Si concentration (2%, 3%, 4%, 5%) under extremely high cooling rate is studied by MD simulations. The atomic structures of the obtained samples are analyzed by cluster alignment method to quantify the SRO and MRO in these systems. It is found that pure Pd undergoes crystallization transition by cooling at the rate of $10^{12}$ K/s. As the cooling rate reaches $10^{13}$ K/s, pure liquid Pd is solidified into amorphous state but there are nearly no glass-forming motifs in the sample. However, the calculated results indicate that a minor amount of Si atoms doping (~4%) could assist Pd-Si system to achieve the glassy state under the cooling rate of $10^{12}$K/s, implying the enhancement of GFA. In the Si-doped Pd glassy samples, one type of dominant glass-forming motif is Pd-centered Z13 clusters which is absent in the amorphous pure Pd sample. Moreover, most of the Si-centered clusters are similar to the glass-forming Z9-like or Mixed-ICO-Cube motif [24]. It is also found that these glass-forming clusters tend to aggregate and form a network structure in the $Pd_{1-x}Si_x$ samples.

II. Method

MD simulations of pure Pd and $Pd_{1-x}Si_x$ intermetallic alloys are performed using LAMMPS software package [25]. The semi-empirical Pd-Si potential based on Embedded Atom Method (EAM) of Finnis–Sinclair (FS) type [26] is used in the simulations. The isothermal-isobaric ensemble (NPT, N=5000 atoms, P=0) and a Nose-Hoover thermostat are utilized throughout the sample preparation. Periodic boundary conditions are used. The time step for integration in the MD simulations is 2.5 fs. First, all the samples are equilibrated at 2000 K for 2 ns and then quenched to 300 K at different cooling rates. To eliminate the effect of atomic thermal motions, the snapshots over 500 ps are collected to calculate the averaged structure of the sample at T=300K. For structural analysis at other temperatures, the inherent structures which are obtained by freezing the sample rapidly are used.

In order to classify the SROs in the MD samples, we employed the cluster alignment method [12] to analyze the similarity between the clusters in the sample and the known templates. The atoms at the first shell around every Pd or Si atoms are extracted from the MD samples and these Pd- and Si-centered clusters are aligned against the templates, e.g., HCP, FCC and BCC, etc. The corresponding alignment score is defined as [12]

$$s = \min_{0.8 \leq \alpha \leq 1.2} \left[\frac{1}{N}\sum_{i=1}^{N}\frac{(\vec{r}_{ic}-\alpha\vec{r}_{it})^2}{(\alpha\vec{r}_{it})^2}\right]^{1/2} \quad (1)$$

where $N$ is the number of the neighbor atoms in the template. $\vec{r}_{ic}$ and $\vec{r}_{it}$ are the atomic positions in the aligned cluster and template, respectively. This score represents the similarity between the aligned cluster and the template, i.e., the smaller the score is, the more similar is the cluster to the template. Here $\alpha$ is chosen between 0.8 and 1.2 to vary the size of the template for an optimal alignment. Thus, for one specific template, we can count the number of clusters with the corresponding alignment score in the interval $[s, s + \Delta s]$ and divide it by total number of clusters, and then obtain the alignment score distribution function $f(s)$. For assigning the clusters to the given motif, an alignment score cutoff of 0.16 is used here, i.e., we count the number of clusters with alignment score $s$ < 0.16 against one specific template to calculate the fraction of SRO. Moreover, if a cluster has alignment scores less than the cutoff for more than one templates, the cluster is assigned to the template with the lowest alignment score.

III.   Results and Discussions

   3.1 Pure Palladium Samples

Comparison of the total structure factor S($q$) from MD simulation with experimental measurement is the most convenient metrics to check whether the semi-empirical potential is good or not to study the atomic structure of liquid and glass samples. Due to the lack of the experimental data of binary $Pd_{1-x}Si_x$ liquids and glasses with mall $x$, we only calculated the S($q$) of pure liquid Pd sample using the semi-empirical Pd-Si EAM potential to compare with S($q$) obtained from the experiment [27]. Fig. 1 shows the calculated and the experimental S($q$) of pure liquid Pd at the temperature of 1853K. One can observe that the total structure factor calculated from our MD simulation agrees well with the experimental data, which is an essential requirement of the reliable MD simulations in the following calculations.

To investigate the phase transition of pure Pd as the function of temperature, the evolution of instantaneous potential energies $E$-3$k_B$T of MD sample for cooling at different rates ($10^{13}$K/s, $10^{12}$K/s, $10^{11}$K/s and $10^{10}$K/s) are shown in Fig. 2(a). For the sample prepared by quenching at cooling rate $\leq 10^{12}$K/s, the instantaneous potential energy has an abrupt drop during the cooling process, implying the occurrence of crystallization. As shown in the inset of Fig. 2(b), the atomic structure of Pd sample at T=300K obtained by a cooling at the rate of $10^{12}$K/s clearly exhibits crystalline character. In comparison, Fig. 2(a) demonstrates that once the cooling rate is increased to $10^{13}$K/s the temperature dependence of energy $E$-3$k_B$T is a continuous curve and only its slope changes upon cooling, implying a transition from liquid to amorphous state. Indeed, the corresponding atomic structure at T=300K shown in the inset of Fig. 2(c) presents amorphous characteristics.

It is well-known that the energy of ICO SRO is lower than that of close-packed HCP or FCC with Lennard-Jones interaction. Many experiments [28-32] and theories [33-38] have demonstrated that there are ICO clusters in the metallic liquids and glass samples. Furthermore, the main structural feature of ICO is the 5-fold local symmetry and it is believed that the glass formation of the metallic material is highly correlated with the local 5-fold symmetry atomic structure. Consequently, the Z-clusters [39] which have 4, 5 and 6-fold local symmetries should be considered in the structural analysis of glassy samples. In order to investigate the SROs in the amorphous Pd samples obtained from the MD simulations and find out whether the amorphous phase is in metallic glassy state or not, we utilized the cluster alignment method to identify the dominant motifs around Pd atoms.

The distributions of alignment scores against the most common templates for the sample with cooling rate of $10^{13}$ K/s are shown in Fig. 2 (c). It is noted that the calculated coordination number of the amorphous Pd sample is around 12.3. Hence, we only pick up the Z13 cluster as the template from all the Z-clusters. Taking alignment score 0.16 as the cutoff to identify motifs, the dominant clusters are HCP (~30.8%), FCC (~23.9%) and intertwined-cube (~13.3%) in the amorphous sample, while the glass forming motifs with 5-fold symmetry have only very small fractions, e.g., ICO (~0.1%) and Z13 (~1.5%). The HCP and FCC motifs belong to the close-packed crystalline structures which can be viewed as the 'genes' of crystalline phases. In fact, as shown in Fig. 2(b), for the crystalline sample with cooling rate of $10^{12}$ K/s the alignment scores against HCP and FCC have a strong peak near score zero, suggesting that the Pd atoms are arranged according to the close-packed structure with some distortions. On the other hand, the intertwined-cube motif with trait of crystal is found in the Ni-Zr system recently [24]. It is composed of two interpenetrating BCC cubes with common (111) plane. Although the number of intertwined-cube motifs is very small in crystalline sample, it can be inferred that the existence of intertwined-cube (~13.3% of total clusters) motifs in the amorphous sample originates from the fact that the intertwined-cube in the liquid state has not enough time to relax due to the extremely fast quench process. Actually, as shown in Fig. 2(d), in the liquid state of pure Pd the fraction of intertwined-cube is about 14.5% of total clusters, which is close to that in amorphous state (~13.3%). Nevertheless, the three dominant clusters (HCP, FCC and intertwined-cube) in amorphous pure Pd are all crystalline motifs and the glass forming motifs (ICO and Z13) fail to dominate the atomic structure of this amorphous sample. Hence, our MD simulations revealed that at least the quench process with cooling rate of $10^{13}$ K/s is unable to vitrify pure Pd metal. In addition, Fig .2(d) also shows that 16.1% of total clusters in the liquid state of pure Pd are Z13 motifs, which could play an important role in the glass formation under the critical cooling rate.

3.2 $Pd_{1-x}Si_x$ Samples

Figs. 3(a)-(d) show the instantaneous potential energies $E$-$3k_B$T for the $Pd_{1-x}Si_x$ samples with different Si-doping concentrations ($x$=0.02, 0.03, 0.04, 0.05). The total pair correlation functions (PCFs) for the corresponding MD samples at T=300K are also shown in Figs. 3(e)-(h). First, similar to the case of pure Pd, no abrupt drop in potential energy is observed for all $Pd_{1-x}Si_x$ samples during the quench process with the cooling rate of $10^{13}$ K/s. Meanwhile, the broad second peaks in total PCFs suggest that under cooling rate of $10^{13}$ K/s the solid $Pd_{1-x}Si_x$ (x=0.02, 0.03, 0.04, 0.05) samples are all in the amorphous state. However, as the cooling rate drops to $10^{12}$ K/s the instantaneous potential energy of $Pd_{0.98}Si_{0.02}$ sample has a sharp decrease beginning at ~1000K upon cooling process and many sharp peaks occurred in total PCFs, indicating the crystallization of $Pd_{0.98}Si_{0.02}$. In contrast, for the samples with Si-doping concentration ≥3% the abrupt drop of energy curve disappeared and the total PCFs show the feature of amorphous phase. Hence, our calculations indicate that even very small amount of Si solutes (~3%) in Pd could suppress the nucleation and crystal growth of Pd under the cooling rate of $10^{12}$ K/s. But when the cooling rate decreases to $10^{11}$ K/s the crystalline phase can still be observed in the $Pd_{1-x}Si_x$ samples with Si-doping concentration of 2% to 4%, as shown in Figs. 3(a)-(c). As the content of Si-doping reaches 5% the results show that $Pd_{0.95}Si_{0.05}$ sample is at amorphous state under the cooling rate of $10^{11}$ K/s. Again, once the cooling rate is sufficiently low (~$10^{10}$ K/s), the $Pd_{0.95}Si_{0.05}$ sample still shows crystallization process upon cooling. These results clearly demonstrate that small amount of Si solutes in Pd can strongly affect the behavior of Pd metal upon cooling process, i.e., the existence of a small number of Si atoms could hinder the crystallization process and drive the $Pd_{1-x}Si_x$ sample into amorphous state.

Next, we study the atomic structure of the amorphous phases in $Pd_{0.95}Si_{0.05}$

sample at T=300K prepared by cooling at the rate of $10^{12}$ K/s. Fig. 4(a) shows the distribution of alignment scores of Pd-centered clusters against several common motifs. The fraction of crystalline HCP, FCC, and intertwined-cube SROs are 18.2%, 6.7%, and 15.6%, respectively. Compared to the case of amorphous pure Pd, it can be seen that in the amorphous $Pd_{0.95}Si_{0.05}$ sample the fraction of HCP clusters decreases from 30.8% to 18.2% and that of FCC clusters drops from 23.9% to 6.7%. However, the fraction of intertwined-cube SRO is nearly unchanged (13.3% in pure Pd and 15.6% in the $Pd_{0.95}Si_{0.05}$ sample). For ICO SRO, the fraction is only 1.4% in our amorphous $Pd_{0.95}Si_{0.05}$ sample, which is similar to the case of pure Pd sample where there are almost no glassy ICO clusters in pure Pd amorphous phase. It's worth noting that the fraction of glass-forming Z13 cluster is 14.2% in amorphous $Pd_{0.95}Si_{0.05}$ sample, which is significantly larger than that in amorphous pure Pd (only 1.5%) and close to that in pure liquid Pd (16.1%). Our calculations indicate that there are considerable glass-forming Z13 motifs although the crystalline SROs still have a large fraction among the Pd-centered clusters in the amorphous Si-doped Pd sample.

From the above results, it can be inferred that the Si atoms play an important role in the amorphous transition in the Si-doped Pd. Thus, we will analyze the Si-centered SROs which is most likely to determine the properties of $Pd_{0.95}Si_{0.05}$ sample. Here the calculated coordination number of Si atoms in amorphous $Pd_{0.95}Si_{0.05}$ sample is about 9.5. We note that the Z9-like motif has been found in $Pd_{82}Si_{18}$ alloy by pair-wise alignment method [23]. Mixed-ICO-Cube motif with coordination number 10 has also been found in $Ni_{64.5}Zr_{35.5}$ alloy recently [24]. Therefore, we use these two motifs as templates to characterize the Si-centered clusters. Fig. 4(c) shows the distribution of alignment scores of Si-centered clusters. It can be observed that the common SROs (such as HCP, FCC, BCC and ICO) have a large portion in the high score region (>0.2). If we set 0.16 as the cutoff value under which the fractions of common motifs (HCP, FCC, BCC and ICO) can be negligible, about 63.6% and 26.0% Si-centered clusters in the sample are Z9-like and Mixed-ICO-Cube motifs, respectively. The fractions of the other motifs are very small (only intertwined-cube has ~1.2%). Therefore, the Z9-like and Mixed-ICO-Cube motifs are the dominant SROs around Si atoms and can characterize the atomic structure of amorphous $Pd_{0.95}Si_{0.05}$ sample.

The Z13, Z9-like and Mixed-ICO-Cube motifs are critical to understanding the atomic structure of Si doped Pd sample because they contain the structural character of glass, i.e., 5-folds symmetry which is popular in ICO SRO of metallic glass. To illustrate this point, the atomic structure of Z13 and Z9-like template is shown in Fig. 4(b) and (d), respectively. In the Z13 cluster, one position has 4-fold symmetry and two positions have 6-fold symmetry. The other 10 positions have 5-fold symmetry, which is responsible for the glass-forming character of this motif. For the Z9-like motif, there are 9 atoms around the center Si atom and our simulations revealed that these 9 atoms are all Pd atoms. Because this cluster is similar to the standard Z9 cluster with the same Voronoi index, we named this motif as 'Z9-like'. The Z9-like cluster has both crystalline and glassy characters. As shown in Fig. 4(d), the atoms Pd2, Pd3, Pd4 and Pd6 formed a nearly square surface. Also, the atoms Pd5, Pd7, Pd8 and Pd9 formed another nearly square surface which rotates about 45 degrees relative to the former square. Moreover, the atom Pd1 just locates at the BCC center of the cubic has one face formed by Pd5, Pd7, Pd8 and Pd9. Thus, the Z9-like cluster has the character of the cubic crystal. On the other hand, the nine Pd atoms surrounding the Si center present the character of ICO cluster, i.e., the atoms Pd5, Pd7, Pd8 and Pd9 with their Pd neighbors have the structure with 5-fold symmetry and they form 12 triangle faces similar to the 20 triangle faces in ICO. Therefore, from this perspective, the Z9-like cluster could be regarded as a mini version of ICO motif with some distortions. From the above structural analysis, it suggests that the Z13 and Z9-like clusters are both glass-forming motifs and the obtained amorphous $Pd_{1-x}Si_x$ sample with the dominant Pd-centered Z13 and Si-centered Z9-like motifs is in metallic glass phase.

To demonstrate the influence of Si solutes on the atomic structure more clearly, the fractions of various SROs in $Pd_{1-x}Si_x$ sample at T=300K under the cooling rate of $10^{12}$ K/s are plotted in Fig. 5. For Pd-centered clusters, with the increasing of the Si-doping concentration $x$, the fractions of crystalline HCP and FCC SROs drop significantly. It is noted that although the fraction of FCC SRO increases at $x$=0.02 the total amount of FCC and HCP SROs is close to the value of pure Pd. On the contrary, the intertwined-cube and Z13 clusters start to grow when $x$>0.02 and fractions reach about 15% respectively when Si concentration $x \geq$ 0.04. Since the intertwined-cube motif can be viewed as the large distorted BCC cubes, the increasing of intertwined-cube could be correlated with the fact that Si solutes would suppress the crystal growth of FCC Pd. On the other hand, it can be seen that most of Si-centered clusters are HCP or FCC when Si concentration is small ($x$=0.02), implying that the doping Si atoms substituted the lattice of Pd atoms in the crystalline sample. Moreover, the fractions of HCP and FCC drop sharply while almost all the Si-centered clusters are Z9-like or Mixed-ICO-Cube motif when $x$=0.04. It clearly indicates that a small amount of Si solutes (~4%) would significantly prevent the crystallization of Pd and allow the glass-forming clusters (Z13, Z9-like and Mixed-ICO-Cube) dominating the SROs in the samples thus enhance the GFA of Pd metal.

From Figs. 3(d) and (h) one can observe that when the cooling rate reduces to $10^{10}$ K/s the $Pd_{0.95}Si_{0.05}$ sample can undergo crystallization even though the Si-doping concentration is 5%. Hence, it is interesting to study the evolution of Si-centered clusters in $Pd_{0.95}Si_{0.05}$ sample with quench process at the cooling rate of $10^{10}$ K/s. Fig. 6(a) shows the temperature dependence of fractions for various SROs from T=1500K to 300K. It can be seen that before solidification the fractions of glass-forming motifs (Z13, Z9-like and Mixed-ICO-Cube) in the liquid sample increase upon the cooling, and between 1000K and 700K these clusters disappear quickly with the decreasing temperature. At the same time, the fraction of HCP and FCC clusters grow rapidly. This result clearly illustrates the competition between the glass-forming clusters (Z13, Z9-like and Mixed-ICO-Cube) and crystalline (HCP and FCC) clusters in $Pd_{1-x}Si_x$ system. As long as the cooling rate is sufficient low, the system would have sufficient time to overcome the crystalline nucleation barrier and then the sample inevitably experiences crystallization process. The Z13, Z9-like and Mixed-ICO-Cube clusters are transformed into HCP or FCC during this process. However, as shown in Fig. 6(b), when the cooling process is too fast ($10^{12}$ K/s in $Pd_{0.95}Si_{0.05}$) so that the glass-forming clusters in the liquid state can survive, the growth of crystalline clusters is suppressed.

The competition between glass-forming motifs and crystalline motifs can also be seen in the cooling rate dependence of SROs, as illustrated in Fig. 7. For $Pd_{0.95}Si_{0.05}$ sample, when the cooling rate $\geq 10^{11}$ K/s the fraction of Si-centered Z9-like motif is saturated while the Pd-centered Z13 and Si-centered Mixed-ICO-Cube clusters continue to grow with the increase of the cooling rate. In contrast, the Si-centered HCP and FCC completely disappeared and the fractions of Pd-centered HCP and FCC continue to decrease with the increase of cooling when the cooling rate $\geq 10^{11}$ K/s. It is interesting to note that the fraction of Pd-centered Intertwined-Cube motif increases to ~15% as cooling rate $\geq 10^{11}$ K/s although this SRO belongs to crystalline motif. With the ultrafast quenching ($\geq 10^{11}$ K/s), the fraction of this motif in solid sample is just close to the value of pure Pd liquid state (~14.5%, as shown in Fig. 2(d)), implying that the Intertwined-Cube clusters in the liquid state survived in the solidification. Moreover, because Intertwined-Cube motif has a structure of two BCC cubes with titled interpenetrating each other, it can be regarded as an incomplete crystalline motif. When cooling rate is reduced (<$10^{11}$ K/s) it has enough time to relax and lower its energy states, i.e., it transforms into other perfect crystalline motif, which could be inferred from the Fig. 7(a).

Finally, although the glass-forming Z13, Z9-like and mixed-ICO-Cube clusters with 5-fold local symmetry are the dominant SROs in the $Pd_{1-x}Si_x$ sample, the packing of these SROs, i.e., medium-range order, is believed to determine the mechanical response of the metallic glass. For example, the well-known Cu-Zr system has excellent strength and elastic limit. The structural analysis [40-42] shows that the ICO clusters in metallic glass Cu-Zr system tend to connect with each other and form a network structure, which enhanced the mechanical properties of materials under deformation. In $Pd_{1-x}Si_x$ system, it is interesting to study the spatial distribution of Pd-centered Z13 and Si-centered Z9-like and Mixed-ICO-Cube clusters. Fig. 8 shows the spatial distribution of Pd-centered Z13 and Si-centered Z9-like and Mixed-ICO-Cube SROs for the samples at 300K with different Si-doping concentration, where the cooling rate is $10^{12}$ K/s. In this figure, if two Z13 centers are within the distance 0.35nm (the first minimum of total PCF), there is a red bond to connect them. The blue atoms are Si centers of Z9-like and Mixed-ICO-Cube clusters. From Figs. 8(a) and (b) it can be seen that the Z13 centers are either isolated or form short string-like structure when the Si concentration is less or equal to 3%. On the contrary, for Si concentration ≥4% the Z13 cluster begin to aggregate and form a dense network in which the clusters penetrate each other. Moreover, the percolation phenomenon also occurrence in the sample with Si concentration ≥4%, implying that the network extends the whole space of sample forming a backbone-like structure. In the meantime, the Si centers of Z9-like and Mixed-ICO-Cube clusters fill the residual space and thus the sample is mainly composed of glass-forming clusters.

The Si-centered Z9-like and mixed-ICO-Cube clusters also form network structure as the Si concentration ≥ 4%, as shown in Fig. 9. Since the distances between Si centers are relatively larger than that between the Pd centers of Z13 clusters, the network developed by Z9-like and mixed-ICO-Cube is slightly open. Besides the Si centers of the Z9-like and mixed-ICO-Cube, in Fig. 9 all the atoms involved in the clusters are plotted and connected by bonds. It can be seen that there's only one Z9-like cluster in $Pd_{0.98}Si_{0.02}$ due to the small amount of Si doping in the sample. When the concentration of Si grows to 3%, the number of Z9-like and Mixed-ICO-Cube clusters increases and these clusters tend to be gathering together to form string-like structures. But the fraction of Z9-like and mixed-ICO-Cube clusters is not sufficient to develop a network structure. Once the content of Si-doping ≥4%, more and more string-like structures begin to be connected and form a network with penetrating clusters. Figs. 9(c) and (d) also show the trend that as the concentration of Si-doping increases the network tend to connect more compactly with each other. Therefore, from the perspective of MRO our calculated results demonstrate that the glass-forming clusters in $Pd_{1-x}Si_x$ sample are arranged as a network, which is another important feature of the metallic glass material.

IV. Summary

In summary, we studied the atomic structure of fast quenched pure Pd and Si doped Pd liquid with different concentrations (Si contents are 2%, 3%, 4%, 5%, respectively) by MD simulations. Cluster alignment method has been employed to analyze the atomic structure of the trajectories obtained by MD simulations. The calculated results show that crystallization will occur in pure Pd even under extreme high cooling rate of $10^{12}$ K/s. Although after solidification the pure Pd sample could be in amorphous state as the cooling rate rises up to $10^{13}$ K/s, the alignment analysis reveals that the dominant SROs are still crystalline clusters instead of glass-forming clusters.

On the contrary, the MD simulations demonstrate that when a small amount of Si atoms (Si concentration ~4%) are doped into Pd, the sample can be in glass state when quenching at a cooling rate faster than $10^{12}$ K/s. In the glass sample the Pd-centered clusters is dominated by crystalline (HCP, FCC and Intertwined-Cube) and glass-forming (Z13) motifs, while most of Si-centered clusters belong to the glass-forming Z9-like or Mixed-ICO-Cube motif. Furthermore, we studied the spatial distribution of the glass-forming clusters. The results show that as more Si atoms are doped into Pd the corresponding Z13, Z9-like and mixed-ICO-Cube clusters tend to aggregate and interpenetrate with each other to form a network throughout the sample.

### Acknowledgements

Work at Ames Laboratory was supported by the U.S. Department of Energy (DOE), Office of Science, Basic Energy Sciences, Materials Science and Engineering Division including a grant of computer time at the National Energy Research Supercomputing Center (NERSC) in Berkeley. Ames Laboratory is operated for the U.S. DOE by Iowa State University under contract # DE-AC02-07CH11358. This work was also supported by the National Natural Science Foundation of China (Grant

Nos. 11304279 and 11104247). Z. J. Yang acknowledges the Natural Science Foundation of Zhejiang Province, China (Grant No. LY18E010007). T. Q. Wen acknowledges the financial support from the National Natural Science Foundation of China (Grant Nos. 51671160 and 51271149).

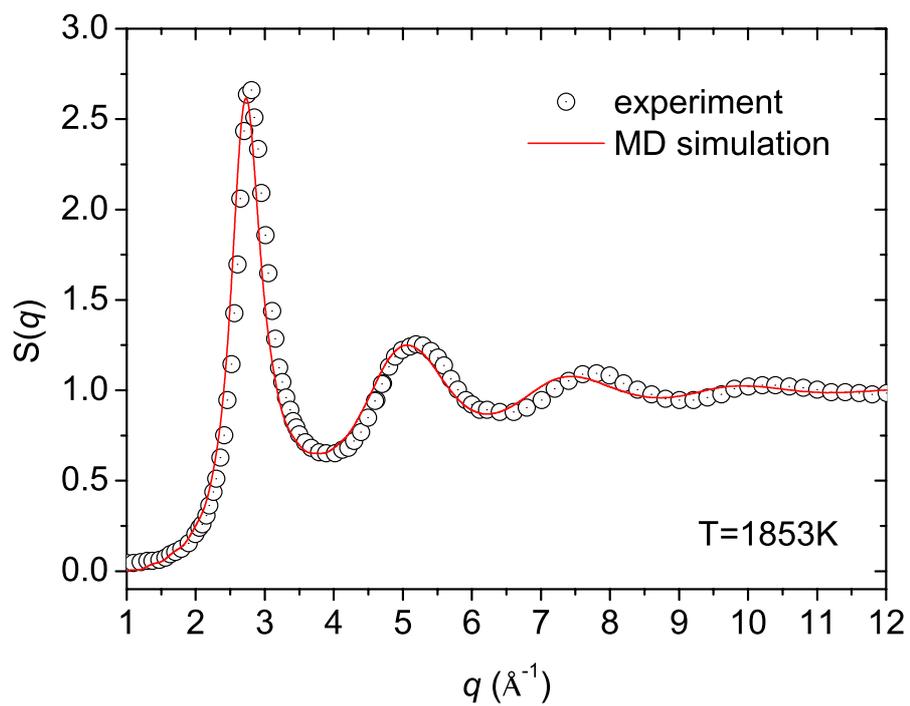

Fig. 1. The comparison of total structure factors for liquid Pd at T=1853K from MD simulation to experimental data. The red line is the result of MD simulation with semi-empirical Pd-Si EAM potential.

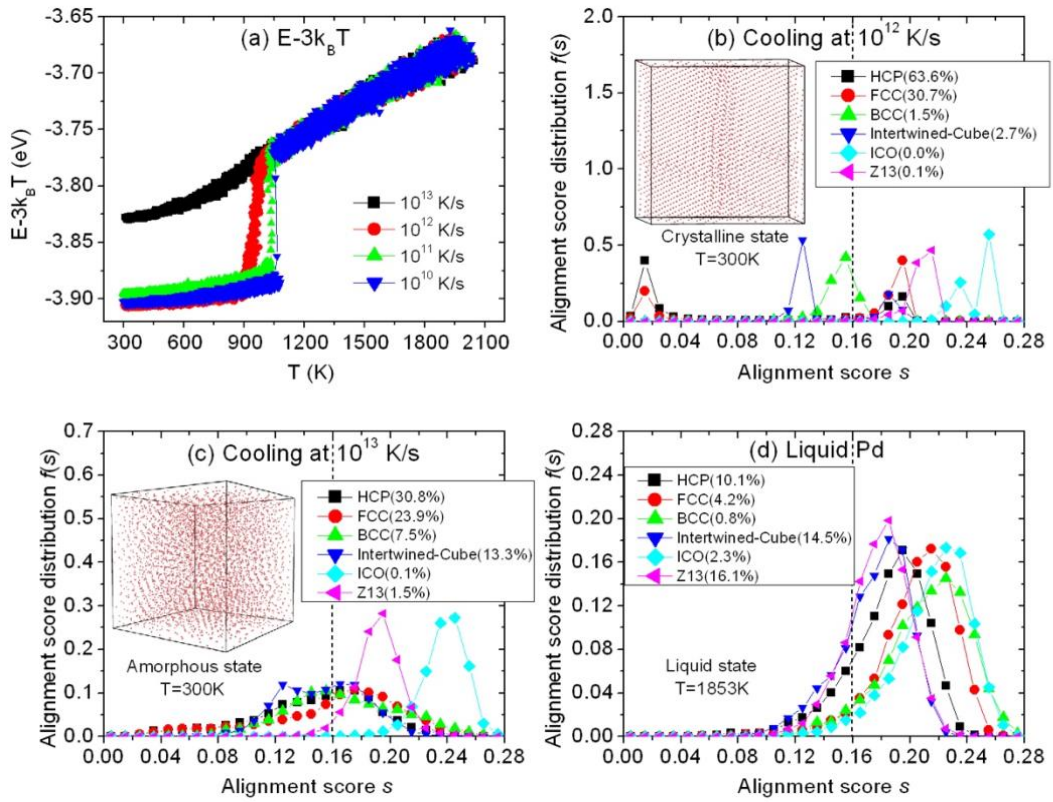

Fig. 2. Phase transition and structural analysis for pure Pd sample. (a) The instantaneous energies $E-3k_BT$ as a function of temperature with different cooling rates. The distributions of alignment scores against various motifs for pure Pd at T=300K under cooling at the rate of (b) $10^{12}$ K/s and (c) $10^{13}$ K/s. The corresponding atomic structures of samples are illustrated in the inset of (b) and (c), respectively. (d) The distributions of alignment scores for liquid pure Pd at T=1853K. The percentage values in the legends of figures are the fractions of various motifs in the sample.

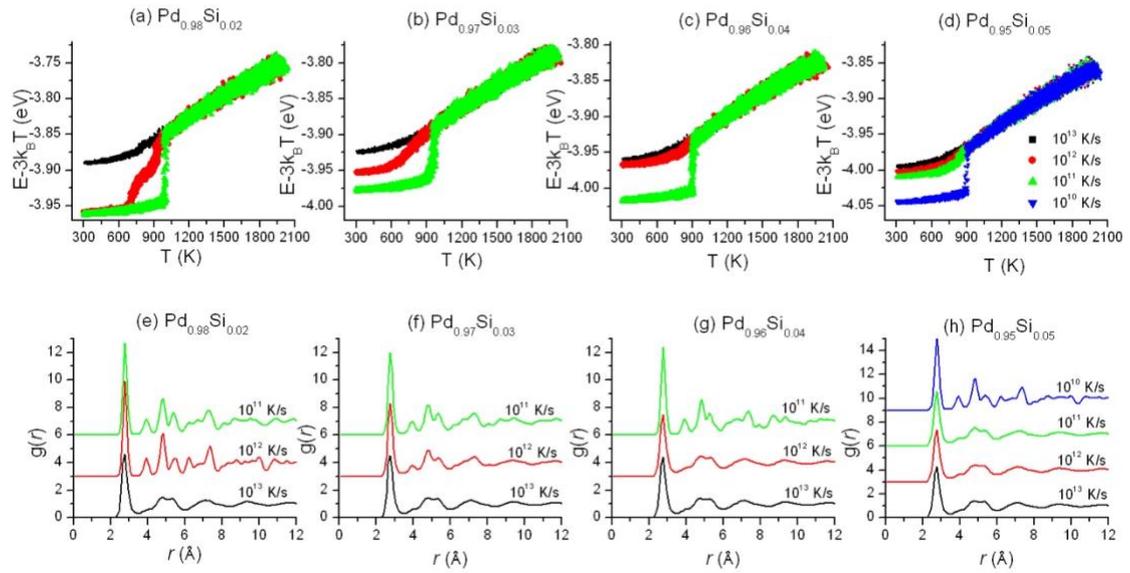

Fig. 3. The temperature dependence of instantaneous energies $E-3k_BT$ with different cooling rates for (a) $Pd_{0.98}Si_{0.02}$, (b) $Pd_{0.97}Si_{0.03}$, (c) $Pd_{0.96}Si_{0.04}$ and (d) $Pd_{0.95}Si_{0.05}$. (e-h) The corresponding total pair correlation functions at T=300K. For the samples with Si-doping concentration $\geqslant 0.03$ under cooling rates $\geqslant 10^{12}$ K/s, the missing of abrupt drop in energy curve and the shape of total $g(r)$ indicate the feature of amorphous phase, i.e., the nucleation and crystal growth of Pd are suppressed by a small amount Si doping.

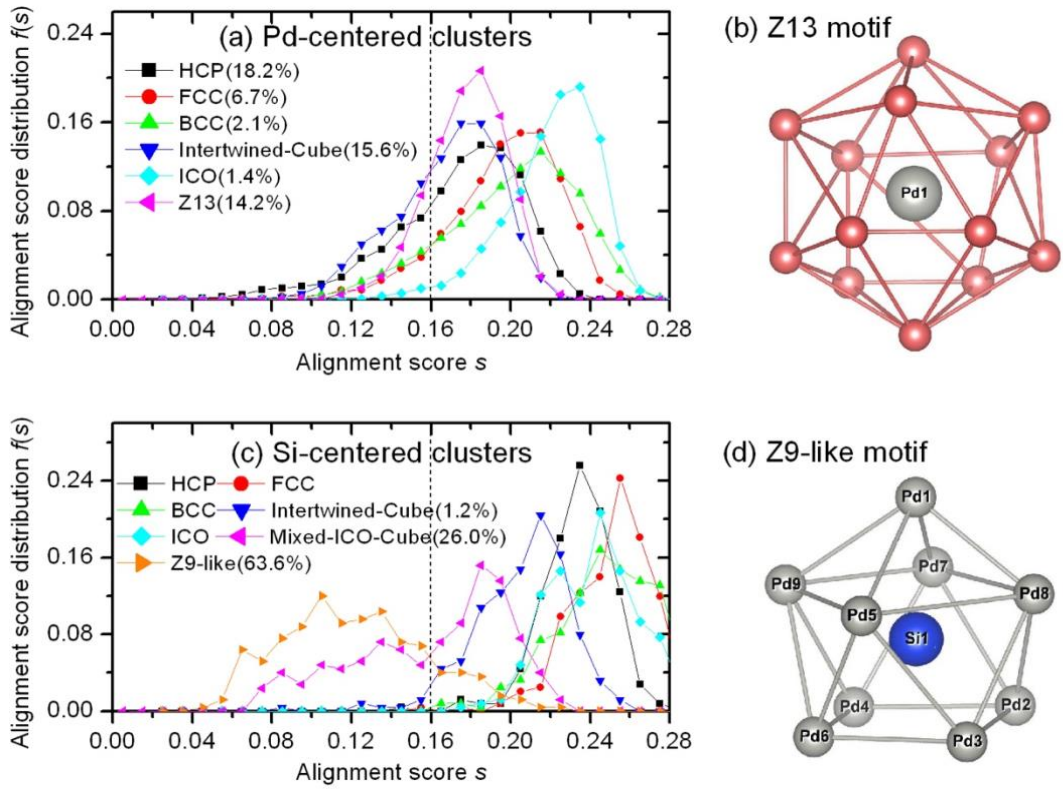

Fig. 4. The structural analysis for $Pd_{0.95}Si_{0.05}$ at T=300K by cooling at the rate of $10^{12}$ K/s. The distributions of alignment scores for (a) Pd- and (c) Si-centered clusters against crystalline (HCP, FCC, BCC and Intertwined-Cube) and glass-forming (ICO, Z13, Mixed-ICO-Cube and Z9-like) motifs. The percentage values in the legends of figures are the fractions of various motifs in the sample. Illustration of glass-forming (b) Pd-centered Z13 and (d) Si-centered Z9-like motifs. Our results indicate that in amorphous $Pd_{0.95}Si_{0.05}$ sample the crystalline SROs still have a large fraction but there are considerable glass-forming Z13 clusters among the Pd-centered clusters. For Si-centered clusters the glass-forming Z9-like and Mixed-ICO-Cube motifs are the dominant SROs.

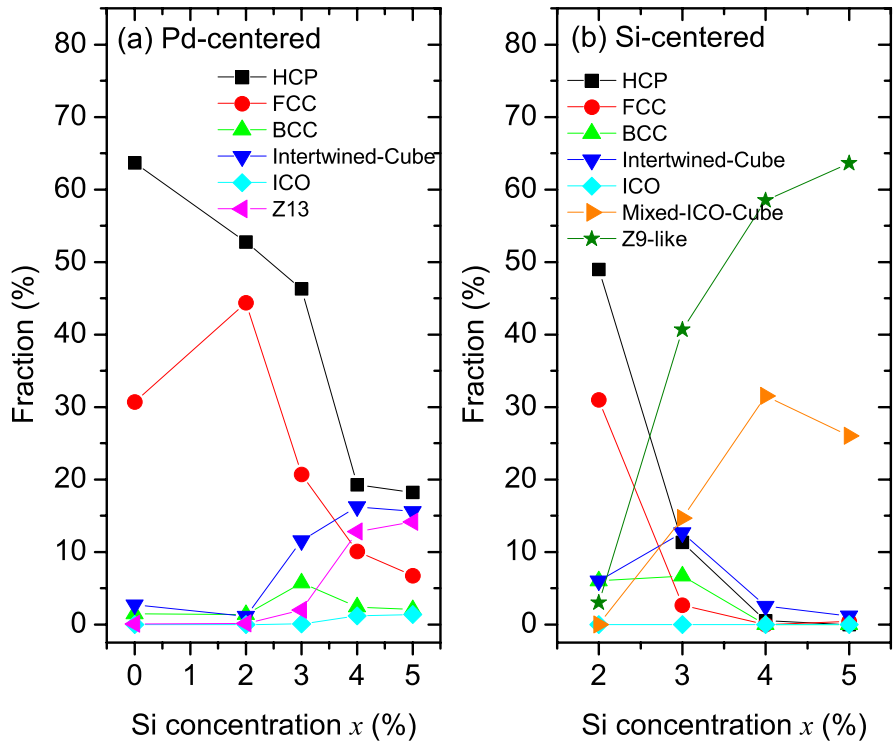

Fig. 5. The fractions of various SROs in Pd$_{1-x}$Si$_x$ at T=300K by cooling at the rate of $10^{12}$ K/s as the function of Si-doping concentration $x$. (a) Pd-centered SROs. (b) Si-centered SROs. It can be seen that a small amount of Si solutes would significantly suppress the crystallization of Pd and increase the fraction of glass-forming SROs (Z13, Z9-like and Mixed-ICO-Cube) in the samples, leading to the enhancement of GFA.

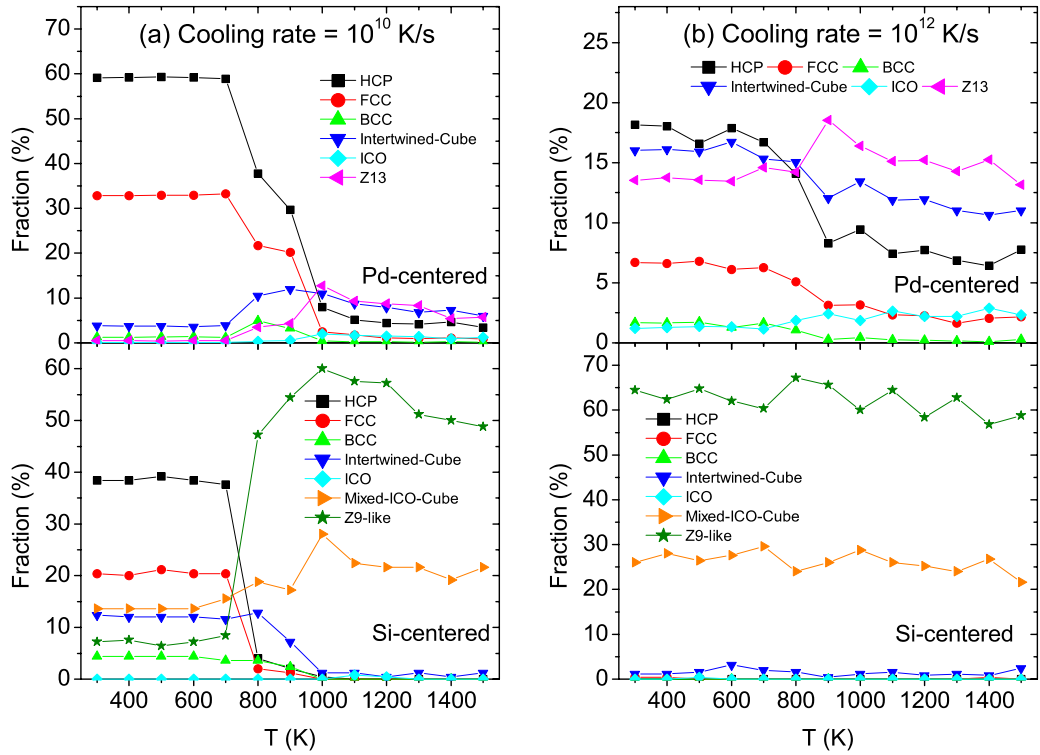

Fig. 6. The evolution of Pd- and Si-centered SROs in $Pd_{0.95}Si_{0.05}$ upon (a) crystalline transition with the cooling rate of $10^{10}$ K/s and (b) glass transition with the cooling rate of $10^{12}$ K/s. Our results imply that under the cooling rate of $10^{10}$ K/s, the Z13, Z9-like and Mixed-ICO-Cube clusters are transformed into HCP or FCC during the crystalline transition. However, when the cooling rate is $10^{12}$ K/s the glass-forming clusters (Z13, Z9-like and mixed-ICO-Cube) in the liquid state can survive in the quench process.

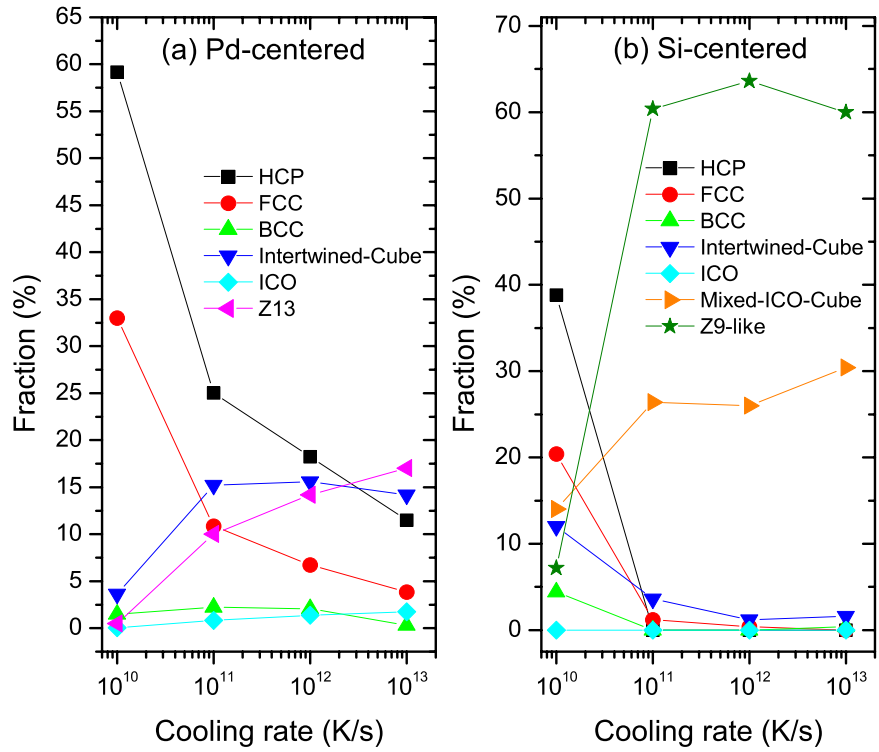

Fig. 7. The fractions of various SROs in $Pd_{0.95}Si_{0.05}$ at T=300K as the function of cooling rate. (a) Pd-centered SROs. (b) Si-centered SROs. Our results clearly exhibit the competition between glass-forming motifs (Z13, Z9-like and Mixed-ICO-Cube) and crystalline motifs (HCP and FCC).

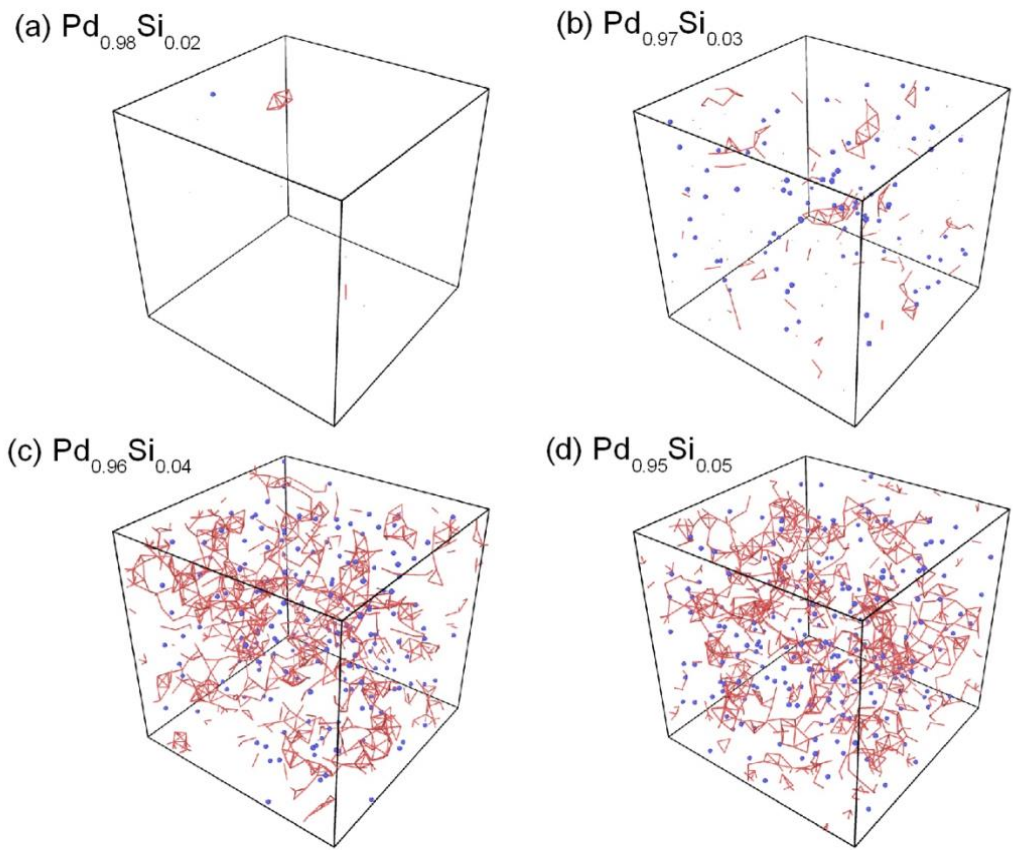

Fig. 8. The spatial distribution of Pd-centered Z13 and Si-centered Z9-like and Mixed-ICO-Cube SROs in (a) $Pd_{0.98}Si_{0.02}$, (b) $Pd_{0.97}Si_{0.03}$, (c) $Pd_{0.96}Si_{0.04}$ and (d) $Pd_{0.95}Si_{0.05}$ at T=300K by cooling under rate of $10^{12}$ K/s. A red bonds is plotted between the Z13 centers with distances shorter than 0.35nm. The blue atoms are Si centers of Z9-like and Mixed-ICO-Cube clusters. It can be seen that for the samples with Si concentration ≥0.04 the Z13 clusters tend to aggregate and penetrate each other, forming a dense network.

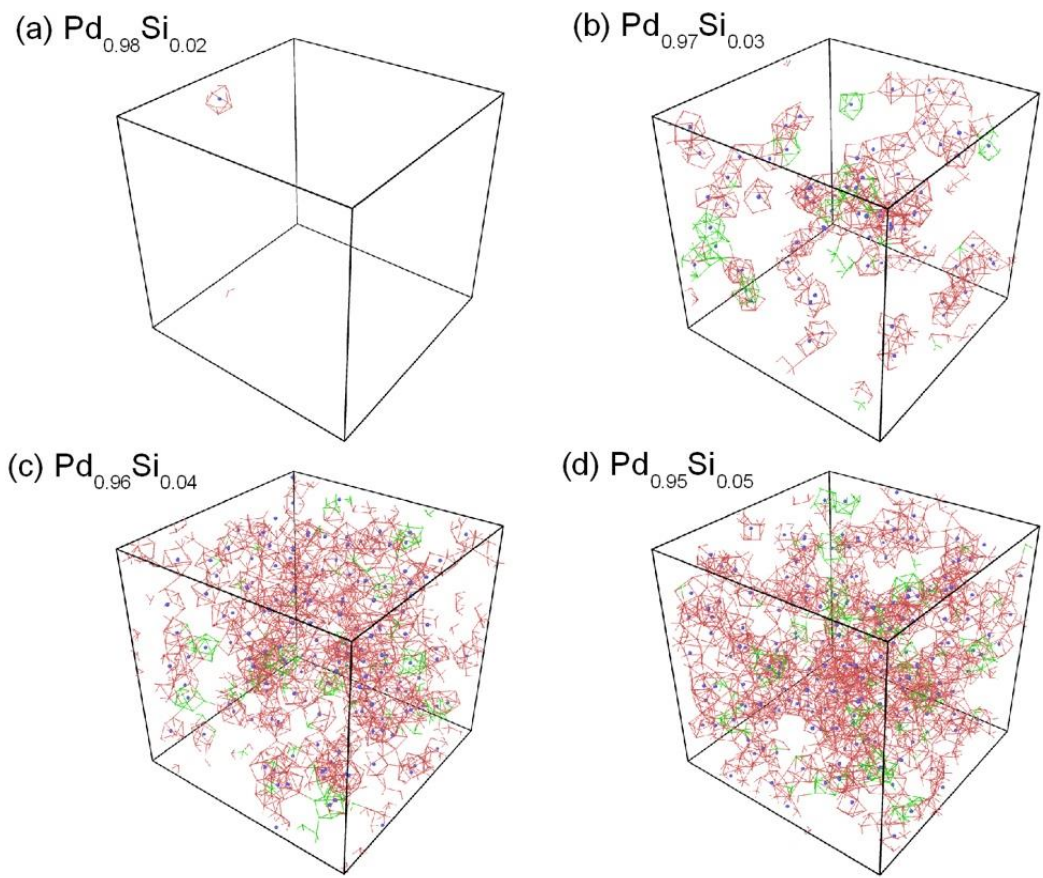

Fig. 9. The spatial distribution of Si-centered Z9-like and Mixed-ICO-Cube SROs in (a) $Pd_{0.98}Si_{0.02}$, (b) $Pd_{0.97}Si_{0.03}$, (c) $Pd_{0.96}Si_{0.04}$ and (d) $Pd_{0.95}Si_{0.05}$ at T=300K by cooling under the rate of $10^{12}$ K/s. The blue atoms are Si centers of Z9-like and Mixed-ICO-Cube clusters, whereas the other atoms involved are connected using red (Pd in Z9-like) and green (Pd in Mixed-ICO-Cube) bonds. It can be seen that as the concentration of Si-doping increases the Si-centered Z9-like and Mixed-ICO-Cube clusters form a network structure and tend to connect more compactly with each other.